\documentclass{elsart}
\usepackage{graphicx}
\usepackage{epsfig}

\def\be{\begin{equation}}
\def\ee{\end{equation}}
\def\bea{\begin{eqnarray}}
\def\eea{\end{eqnarray}}

\def\NN{{\mathcal V}}

\begin{document}

\begin{frontmatter}

\title{\bf Simple solutions of relativistic hydrodynamics\\
	for longitudinally expanding systems}

\author[KFKI,USP]{T. Cs\"org\H o,\thanksref{tamas}}
\author[USP]{F. Grassi,\thanksref{frederique}}
\author[USP]{Y. Hama,\thanksref{yogiro} and}
\author[UFRJ]{T.  Kodama\thanksref{takeshi}}
\address[KFKI]{MTA KFKI RMKI, H - 1525 Budapest 114, POBox 49, Hungary}
\address[USP]{IF, USP, C. P. 66318, 05389-970 S\~{a}o Paulo, SP, Brazil}
\address[UFRJ]{IF, UFRJ, C. P. 68528, 21945-970 Rio de Janeiro, RJ, Brazil}
\thanks[tamas]{Email: csorgo@sunserv.kfki.hu}
\thanks[frederique]{\phantom{Email:} grassi@fma.if.usp.br}
\thanks[yogiro]{\phantom{Email:} hama@fma.if.usp.br}
\thanks[takeshi]{\phantom{Email:} tkodama@if.ufrj.br}

\date{Febr. 8. 2002}

\begin{abstract}
Simple, self-similar, analytic solutions of 1 + 1 dimensional relativistic
hydrodynamics are presented, generalizing Bjorken's solution to inhomogeneous
rapidity distribution.
\end{abstract}
\end{frontmatter}

\section{Introduction}

Relativistic hydrodynamics has various applications, including the
calculations of single-particle spectra and two-particle correlations in high
energy heavy ion collisions. The applications of relativistic hydrodynamics in
the field of heavy ion physics have been reviewed in ~\cite{csernai}. More
recently, there has been an increasing interest in studying RHIC and coming
LHC experiments in the framework of relativistic hydrodynamics 
\cite{Shu,Hirano,Kolb}. 
The hydrodynamical analysis can also be extended to the study
of these processes on event-by-event basis \cite{SPH-qm01,SPH-jpg}. However,
most works in hydrodynamics are numerical so not always transparent. In this
sense, exact solutions would be useful, but are rarely found due to the highly
non-linear nature of relativistic hydrodynamics. Actually, Landau's
one-dimensional analytical solution of relativistic hydrodynamics
\cite{Landau} gave rise to a new approach in high energy physics. The
boost-invariant Bjorken solution \cite{Bjorken}, found more than 20 years
later, is frequently utilized as the basis for estimations of initial energy
densities in ultra-relativistic nucleus-nucleus collisions. Recently, Bir\'{o}
has found  self-similar exact solutions of relativistic hydrodynamics for
cylindrically expanding systems \cite{biro1,biro2}, however, his solutions are
valid only when the pressure is independent of space and time, as e.g. in the
case of a rehadronization phase transition in the middle of a relativistic
heavy ion collision.

Here we present an analytic approach, which goes back to the data-motivated
exact analytic solution of non-relativistic hydrodynamics found by
Zim\'{a}nyi, Bondorf and Garpman (ZBG) in 1978 for low energy heavy ion
collisions with spherical symmetry~\cite{jnr}. This solution has been extended
to the case of elliptic symmetry by Zim\'{a}nyi and collaborators in
~\cite{jde}. In ~\cite{nr,nrt} a Gaussian parameterization has been introduced to
describe the mass dependence of the effective temperature and the radius
parameters of the two-particle Bose-Einstein correlation functions in high
energy heavy ion collisions. Later it has been realized that this
phenomenological \textit{parameterization} of data corresponds to an exact,
Gaussian \textit{solution} of non-relativistic hydrodynamics with spherical
symmetry~\cite{cspeter}. The spherically symmetric self-similar solutions of
non-relativistic hydrodynamics were obtained in a general manner in
~\cite{cssol}, that included an arbitrary scaling function for the temperature
profile, and expressed the density distribution in terms of the temperature
profile function. The ZBG solution and the Gaussian solution of
~\cite{cspeter} are recovered from the general solution of ~\cite{cssol} as
special cases, corresponding to different scaling functions of the temperature
profile. The Gaussian solution has been generalized to ellipsoidal expansions in
~\cite{ellsol}, that provides analytic insight into the physics of non-central
heavy ion collisions~\cite{ellsp}.

Our approach corresponds to a relativistic generalization of these recently
obtained analytic solutions~\cite{cspeter,cssol,ellsp} of non-relativistic
fireball hydrodynamics to the case of relativistic flows, based on the success
of the analytic approach to parameterize the single particle spectra and the
two-particle Bose-Einstein correlations in high energy heavy ion physics in
terms of hydrodynamically expanding sources~\cite{3d}.

In particular, we attempt here to solve the 1+1 dimensional relativistic
hydrodynamical problem, in trying to overcome two shortcomings of Bjorken's
well-known solution. These two shortcomings of Bjorken's solutions are that
\textit{i)} it only describes ultra-relativistic limit so that the rapidity
distribution is flat; \textit{ii)} it contains no transverse flow.

Here we present a new family of exact analytic solutions of relativistic
hydrodynamics in 1 + 1 dimension (time + longitudinal coordinate), that is
able to describe arbitrary inhomogeneous rapidity distributions. The inclusion
of the transverse flow will be treated in a subsequent paper.

An interesting aspect of this solution is that the shapes of the rapidity
distribution $dN/dy$ and temperature distribution are coupled in the way that
the larger the rapidity density, the smaller the effective temperature.
Choosing the effective temperature distribution $T_{\mathrm{eff}}(y)$ to be
flat, we recover Bjorken's 1+1 dimensional solution.

\section{The equations of relativistic hydrodynamics}

In order to get solutions that have simple non-relativistic limiting behavior,
we consider a gas of one type of particles whose total number is conserved. We
then solve the relativistic continuity and energy-momentum conservation
equation:
\begin{eqnarray}
\partial_{\mu}\left(  nu^{\mu}\right)    & = & 0,\label{e:cont}\\
\partial_{\mu}T^{\mu\nu}  & = & 0.
\end{eqnarray}
Here $n\equiv n\left(  x\right)  $ is the number density, the four-velocity is
denoted by $u^{\mu}\equiv u^{\mu}\left(  x\right)  =\gamma\left(
1,\mathbf{v}\right)  $, normalized to $u^{\mu}u_{\mu}=1$, and the
energy-momentum tensor is denoted by $T^{\mu\nu}$. We assume perfect fluid,
\begin{equation}
T^{\mu\nu}=\left(  \varepsilon+p\right)  u^{\mu}u^{\nu}-pg^{\mu\nu},
\end{equation}
where $\varepsilon$ stands for the relativistic energy density and $p$ denotes
the pressure.

We close this set of relativistic hydrodynamical equations with the use of an
equation of state. We assume an ideal gas that contains massive conserved
quanta,
\begin{eqnarray}
\varepsilon & = & mn+\kappa p,\label{e:eos1}\\
p  & = & nT.\label{e:eos2}%
\end{eqnarray}
This equation of state has two free parameters, $m$ and $\kappa$. The case of
ultrarelativistic gas is recovered by choosing $m=0$ and $\kappa=3$. If the
thermal motion of the gas is non-relativistic, one has $m\gg T$. Using this
condition, the equations of non-relativistic hydrodynamics are re-obtained
from the above set of equations in the $\mathbf{v}^{2}\ll1$ limiting case. The
non-relativistic ideal gas equation of state corresponds to the case of
$\kappa=3/2$.

The energy-momentum conservation equation contains 4 independent equations.
These can be projected into a component parallel to $u^{\mu}$ and a component
orthogonal to $u^{\mu}$, that yield the relativistic energy and Euler
equations:
\begin{eqnarray}
u^{\mu}\partial_{\mu}\epsilon+ (\epsilon+ p) \partial_{\mu}u^{\mu} &  = &
0,\label{e:ren}\\
u_{\nu}u^{\mu}\partial_{\mu}p + (\epsilon+ p)u^{\mu}\partial_{\mu}u_{\nu}-
\partial_{\nu}p  &  =  & 0.\label{e:rEu}%
\end{eqnarray}
The relativistic Euler equation contains only three independent components, by construction.

Using the equation of state and the continuity equation, the energy equation
can be rewritten as an equation for the temperature,
\begin{equation}
u^{\mu}\partial_{\mu}T + \frac{1}{\kappa} T \partial_{\mu}u^{\mu}=
0.\label{e:rT}%
\end{equation}

From now on we assume that the flow is homogeneous in the transverse direction
corresponding to infinitely broad target and projectile in high energy heavy
ion collisions. We reduce our equations to the 1 + 1 dimensional case, so the
coordinates are $x^{\mu}= (t, r_{z})$, $x_{\mu}= (t, -r_{z})$ and the metric
tensor is $g^{\mu\nu} = g_{\mu\nu} = \mbox{\rm diag}(1,-1)$.

We solve 3 independent equations, the continuity, the temperature equation and
the $z$ component of the Euler equation~(\ref{e:cont},\ref{e:rEu},\ref{e:rT}).
The equations, (\ref{e:eos1},\ref{e:eos2}) close this system of equations in
terms of 3 variables, $n$, $T$ and $v_{z}$.

\section{Self-similar solution}

We search for solutions that scale in the $z$ direction. We introduce the time
dependent scaling variable
\begin{equation}
x=\frac{r_{z}^{2}}{Z\left(  t\right)  ^{2}},
\end{equation}
and assume that the longitudinal motion corresponds to a Hubble type of
self-similar longitudinal expansion,
\begin{equation}
v_{z}(t,r_{z})=\frac{\dot{Z}(t)}{Z(t)}r_{z},\label{e:vz}%
\end{equation}
where $\dot{Z}=dZ(t)/dt$ stands for the time derivative of the scale parameter
$Z$. In a relativistic notation, this form is equivalent to
\begin{eqnarray}
u^{\mu}  & = & (\cosh\zeta,\sinh\zeta),\\
\tanh\zeta & = & \frac{\dot{Z}(t)}{Z(t)}r_{z},\\
\cosh\zeta & = & \frac{1}{\sqrt{1-\dot{Z}^{2}x}}\equiv\gamma.
\end{eqnarray}
Using this ansatz, we find that the continuity equation is solved by the form
\begin{equation}
n(t,r_{z})=n_{0}\frac{Z_{0}}{Z}\frac{1}{\cosh\zeta}\mathcal{G}%
(x),\label{s:cont}%
\end{equation}
where $\mathcal{G}(x)$ is an arbitrary non-negative function of the scaling
variable $x$ and $n_{0}$ and $Z_{0}$ are normalization constants. We use the
convention $Z_{0}=Z(t_{0})$ and $n_{0}=n(t_{0},0)$ which implies that
$\mathcal{G}(x=0)=1$. The temperature equation, (\ref{e:rT}) is solved by the
following form:
\begin{equation}
T(t,r_{z})=T_{0}\left(  \frac{Z_{0}}{Z}\frac{1}{\cosh\zeta}\right)
^{1/\kappa}\mathcal{F}(x).
\end{equation}
The constants of normalization are chosen such that $T_{0}=T(t_{0},0)$ and
$\mathcal{F}(0)=1$. Here again, we find that the solution is independent of
the form of the function $\mathcal{F}(x)$, provided that $\mathcal{F}(x)\geq0$.

Using the ansatz for the flow profile and the solution for the density and the
temperature, the relativistic Euler equation reduces to a complicated
non-linear equation that contains $Z$, $\dot{Z}$ and $\ddot{Z}$ and $x$.
Taking this equation at $x=0$ we express $\ddot{Z}$ as a function of $Z$ and
$\dot{Z}$. Substituting this back to the Euler equation we obtain an equation
for $\dot{Z},Z$ and $x$. In particular, for the $m=0$ case, $Z$ cancels out
and this reduces to a second order polynomial equation for $\dot{Z}^{2}$,
which has only one positive root. The form of the solution in this case
($m=0$) is
\begin{equation}
\dot{Z}^{2}(t)=F(x).
\end{equation}
Observing that the function $F$ depends only on the scaling variable $x$,
while $\dot{Z}$ depends only on the time variable $t$, we conclude that the
only solution of this equation should be a constant
\begin{equation}
\dot{Z}=\dot{Z}_{0}.
\end{equation}
Now we choose the origin of the time axis such that $Z(t=0)=0$ without loss of
generality. The solutions can be cast in a relatively simple form by
introducing the longitudinal proper time $\tau$ and the space-time rapidity
$\eta$ defined as
\begin{eqnarray}
\tau & = & \sqrt{t^{2}-r_{z}^{2}},\label{e:tau}\\
\eta & = & 
	\frac{1}{2}\log\left(  \frac{t+r_{z}}{t-r_{z}}\right)  .\label{e:eta}%
\end{eqnarray}
This implies that
\begin{eqnarray}
Z(t)  & = & \dot{Z}_{0}t,\label{e:zsol}\\
v_{z}  & = & \frac{r_{z}}{t}\,=\,\tanh\eta,\label{e:vzsol}\\
\zeta & = & \eta.
\end{eqnarray}
Thus the solution for the flow velocity field corresponds to the flow field of
the Bjorken solution. However, in the Bjorken solution the temperature
distribution was independent of the $\eta$ variable, while in our case the
density and the temperature distributions can be both $\eta$ dependent,
or in other words, our solutions are scale dependent. 
The scale is defined by the 
parameter $\dot{Z}_{0}$, in the longitudinal direction.

This special form of the solution for the flow velocity field implies that
$\ddot Z = 0$. This equation implies that there is no pressure gradient and
there is no acceleration in this class of self-similar solutions. The Euler
equation is reduced to the following requirement:
\begin{equation}
(\partial_{z} + \frac{r_{z}}{t} \partial_{t}) \left[  \left( \frac{ t_{0}
}{\tau}\right) ^{(1+1/ \kappa)} ( 1 - \dot Z_{0}^{2} x)^{(1+1/\kappa)}
\mathcal{G}(x) \mathcal{F}(x) \right]  = 0
\end{equation}
This equation is solved by the trivial $\mathcal{G}(x) \mathcal{F}(x) = 0$ as
well as by the non-trivial solution of
\begin{equation}
\mathcal{G}(x) \mathcal{F}(x) = (1 - \dot Z_{0}^{2} x)^{- (1 + 1/\kappa
)},\label{e:nt}%
\end{equation}
which is indeed only a function of $x$ as $\dot Z_{0}$ is a constant of time.
With this form, the Euler equation is satisfied. This solution implies that
the scaling profile functions for the temperature and the density distribution
are not independent. As the constraint is given only for their product, one of
them can be still chosen in an arbitrary manner.

At this point it is worthwhile to introduce new forms of the scaling
functions. Let us define
\begin{eqnarray}
\mathcal{T}(x)  & = & \mathcal{F}(x)(1-\dot{Z}_{0}^{2}x)^{1/\kappa},\\
\mathcal{V}(x)  & = & \mathcal{G}(x)(1-\dot{Z}_{0}^{2}x).
\end{eqnarray}
Then the constraint Eq.(\ref{e:nt}) can be cast to the simplest form of
\begin{equation}
\mathcal{V}(x)\mathcal{T}(x)=1.
\end{equation}

Let us summarize our new family of solutions of the 1+1 dimensional
relativistic hydrodynamics by substituting the results in the density,
temperature and pressure profiles.

We obtain
\begin{eqnarray}
v_{z}  & = &\frac{r_{z}}{t}=\tanh\eta,\\
x      & = & 
	\frac{r_{z}^{2}}{\dot{Z}_{0}^{2}t^{2}}
	\,=\,\frac{\tanh^{2}\eta}{\dot {Z}_{0}^{2}},\\
n(t,r_{z})  & = & n_{0}\frac{t_{0}}{\tau}\mathcal{V}(x),\label{e:nsol}\\
p(t,r_{z})  & = & p_{0}\left(  \frac{t_{0}}{\tau}\right)  ^{1+1/\kappa},\\
T(t,r_{z})  & = & T_{0}\left(  \frac{t_{0}}{\tau}\right)  ^{1/\kappa}\frac
{1}{\mathcal{V}(x)},\label{e:tsol}%
\end{eqnarray}
where $p_{0}=n_{0}T_{0}$.

This implies that we have generated a new family of exact solutions of
relativistic hydrodynamics: a new hydrodynamical solution is assigned to each
non-negative function $\mathcal{T}(x)$. It can be checked that the above
solutions are valid also for massive particles, the form of the solution is
independent of the value of the mass $m$. The form of solutions depends
parametrically on $\kappa$, that characterizes the equation of state.

\section{Analysis of the solutions}

We have obtained new solutions of the (1+1) dimensional relativistic
hydrodynamical equations which describe a self-similar, streaming flow. The
pressure and the flow profiles are the same as in the 1+1 dimensional Bjorken
solution. In the case of $\mathcal{V}(x)=1,$ we recover Bjorken's solution. In
this limiting case, the pressure, the density and the temperature profiles
depend only on the longitudinal proper time $\tau$.

In the general case, our solution contains a characteristic scale defining
parameter in the longitudinal direction, $\dot{Z}_{0}$, and an arbitrary
scaling function $\mathcal{V}(x)$. Thus we have an infinitely rich new family
of solutions. Let us try to determine the physical meaning of the scaling
function $\mathcal{V}(x)$.

In order to do this we evaluate the single particle spectra corresponding to
the new solutions. Here we neglect any possible dynamics in the transverse
directions, as usual in case of applications of Bjorken's solution. The
four-velocity field of our solutions thus becomes $u^{\mu}=(\cosh
\eta,0,0,\sinh\eta)$. The four-momentum of the observed particles with mass
$m$ is denoted by $k^{\mu}=(m_{t}\cosh$ $y,k_{x},k_{y},m_{t}\sinh$ $y)$. Let us assume
that particles freeze out at a constant longitudinal proper-time $\tau_{f}$,
for the sake of simplicity. This implies freeze-out at a constant pressure,
but at a space-time rapidity dependent temperature and density, and makes it
possible to continue the calculation analytically. The source function of
locally thermalized relativistically flowing particles in a Boltzmann
approximation can be written as
\begin{equation}
S(x,\mathbf{k}) =C({\eta})\, m_{t}\cosh(\eta-y) \, 
	n(x)\, \exp\left( -k^{\mu}u_{\mu }/T\right)  \, \delta(\tau-\tau_{f}),
\end{equation}
where $C(\eta)$ is an $\eta$ dependent normalization factor, 
given by the condition that 
$\int d\mathbf{k}/E\, S(x,\mathbf{k})=n(x)\delta(\tau-\tau_{f})$, 
which implies that
\be
C(\eta) = \left\{4\pi m^2 T(\tau_f,\eta) K_2[m/T(\tau_f,\eta)]\right\}^{-1},
\ee
where 
$ K_\nu(z) = \int_0^\infty dz \exp(-z \cosh t) \cosh(\nu t)$
is the modified Bessel function of the second kind.
\bigskip

The single particle spectrum can be calculated from the emission function as
\begin{equation}
E\frac{d^{3}N}{d\mathbf{k}}=\int\tau d\tau d\eta S(x,\mathbf{k}).
\end{equation}

Substituting our family of new solutions, 
and using $\mathcal{T}(x)=1/\mathcal{V}(x)$, 
we obtain the following form
\begin{eqnarray}
S(x,\mathbf{k})  & = & \, C(\eta) \, m_{t}\cosh(\eta-y) \, n(x) 
f_B(x,\mathbf{k}) \\
f_B(x,\mathbf{k}) & = & 
\exp\left[  {-\frac{m_{t}\cosh(\eta-y)}{T_{0}}}\left(  {\frac{\tau}{t_{0}}
}\right)^{1/\kappa}{\mathcal{V}(\frac{\tanh^{2}\eta}{\dot{Z}_{0}^{2}}
)}\right]  \delta(\tau-\tau_{f}).
\end{eqnarray}
We are interested in the coupling between the measurable rapidity distribution
and the rapidity dependence of the effective temperature in the transverse
directions as obtained from our new family of solutions. 
We assume that $\mathcal{V}(x)$ is a slowly varying function,
 i.e. $d\log\mathcal{V}%
(x)/dx\ll1$ in the region of interest. This assumption implies that the point
of maximal emissivity is located at $\overline{\eta}=y$ with correction terms
of $\mathcal{O}(d\log\mathcal{V}(x)/dx)$
The measurable single-particle
spectra can be written as
\begin{eqnarray}
E\frac{d^{3}N}{d\mathbf{k}}  & = & 2 C(y) \, n_{0}t_{0}\, \mathcal{V}\left(
\frac{\tanh^{2}y}{\dot{Z}_{0}^{2}}\right)  
K_1[m_{t}/T_{\mathrm{eff}}\left(  y\right) ],\\
\frac{dN}{dy}  & = & n_{0}t_{0} \mathcal{V}\left(
\frac{\tanh^{2}y}{\dot{Z}_{0}^{2}}\right).
\label{e:dndy}
\end{eqnarray}
where\\
\begin{equation}
T_{\mathrm{eff}}(y)=\frac{1}{\mathcal{V}\left(  \frac{\tanh^{2}y}{\dot{Z_0}^{2}%
}\right)  }T_{0}\left(  \frac{t_{0}}{\tau_{f}}\right)  ^{1/\kappa}.
\label{e:teffy}
\end{equation}

Note that the $\mathcal{V}$ function is a free fit function that describes the
measurable rapidity distribution, including characteristic scales of the size
of $\dot{Z}_{0}$.

We see that the slope parameter for transverse mass distribution
$T_{\mathrm{eff}}$ is related to the rapidity distribution as%
\begin{equation}
T_{\mathrm{eff}}(y)=T_{0}\left(  \frac{t_{0}}{\tau_{f}}\right)  ^{1/\kappa
}{\frac{dN/dy\left(y=0\right)}{dN/dy}}.
\end{equation}

Figures  1 and 2 illustrate the calculated behavior of the effective temperature
distribution as a function of rapidity for a single Gaussian-like  
and  a double
Gaussian-like ansatz for the measurable rapidity distribution. In case of a
homogeneous rapidity distribution, $dN/dy=C$ we recover Bjorken's result that
the effective temperature distribution is rapidity independent. This behavior
is expected to appear in high energy heavy ion collisions in the infinite
bombarding energy limit.

\section{Summary}

We have found a new family of solutions of (1+1) dimensional relativistic
hydrodynamics. This family solves the continuity equation and the conservation
of the energy - momentum tensor of a perfect fluid, assuming an ideal gas
equation of state. 
%The mass of the particles $m$ and the constant of
%proportionality between the kinetic energy density and the pressure, $\kappa$,
%are free parameters of the solution. 
The flow field coincides with that of
Bjorken's solution. However, the shape of the measurable rapidity
distribution, $dN/dy$ plays the role of an arbitrary scaling function in our
solution, and we obtain that the effective temperature of the transverse
momentum distribution becomes rapidity dependent. Assuming that $dN/dy$ is a
slowly varying function of the rapidity $y$, we find that the effective
temperature is proportional to the inverse of the rapidity distribution,
$T_{\mathrm{eff}}(y)\propto(dN/dy)^{-1}$.

As compared to the well-known Bjorken's solution, we have solved one more
equation, the continuity equation. We have considered equations of state that
have two free parameters, the mass $m$ and $\kappa=\partial\epsilon/\partial
p$. Interestingly, these generalizations resulted in \textit{additional
freedom} in the solution. Our solution, similarly to Bjorken's case, describes
scaling longitudinal flow and a pressure distribution that depends only on the
longitudinal proper time. However, in our case, the pressure is a product of
the local number density and the local temperature, hence one of these can be
chosen in an arbitrary manner. In principle, we obtained that the measured
single particle rapidity distribution can be arbitrary, and this measurable
distribution plays the role of a \textit{scaling function} of both the density
and the inverse temperature in this family of new solutions of relativistic
hydrodynamics. A generalization of the present one dimensional solution to 3
dimensional case can also be done in a similar manner to the nonrelativistic
case\thinspace\cite{ellsol}.

\textbf{Acknowledgments:} T. Cs. would like to thank L. P. Csernai, B.
Luk\'{a}cs and J. Zim\'{a}nyi for inspiring discussions during the initial
phase of this work, and to Y. Hama and G. Krein for kind hospitality during
his stay at USP and IFT, Sao Paulo, Brazil. This work has been supported by
the OTKA grant T026435 of Hungary, the NWO - OTKA grant
N 25487 of The Netherlands and Hungary, and the grants FAPESP 00/04422-7,
99/09113-3, PRONEX 41.96.0886.00, FAPERJ E-26/150.942/99, and CNPq, Brazil.

\vfill\eject

\begin{figure}[tbp]
%\null
\vspace{6.8cm}
\includegraphics{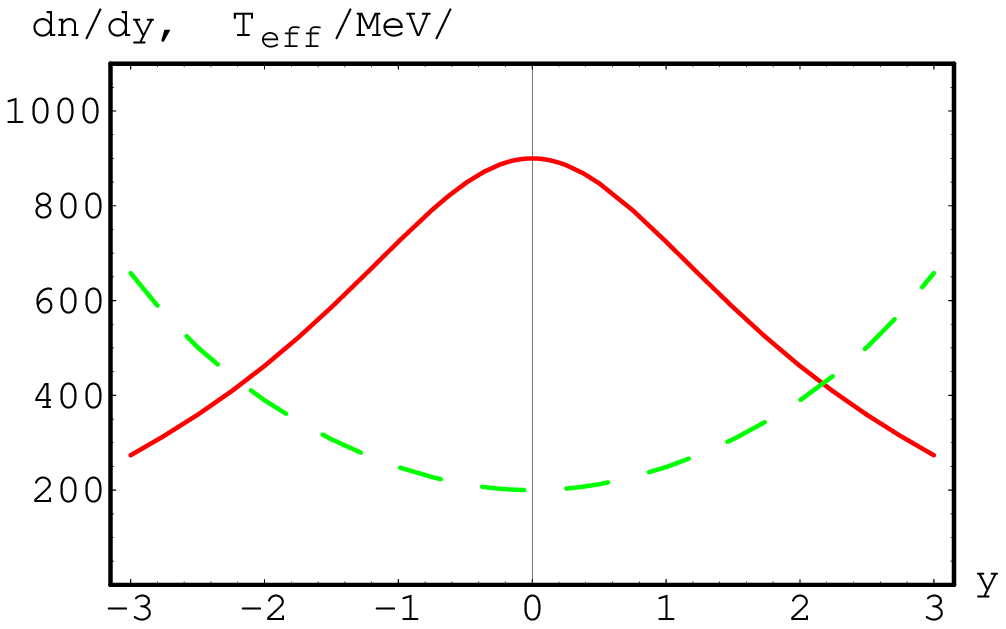}
\vspace{-1.3cm}
\caption{
Rapidity distribution $dN/dy$ and effective temperature distribution
$T_{\rm eff}(y)$ as a function of rapidity $y$, as obtained  from
a new family of solutions of (1+1) dimensional relativistic
hydrodynamics. Here we use the scaling function
$\NN(x) = (1 - x)^{(1/4)}$, using a scale parameter $\dot{Z_0} =\tanh(4)$,
$n_0 t_0 = 900$ and $T_0 (t_0/\tau_f)^{1/\kappa} = 200$ MeV, corresponding to a single maximum in
the rapidity distribution $dN/dy$. 
The analytic expressions are given by eqs.~(\ref{e:nsol},
\ref{e:tsol},\ref{e:dndy},\ref{e:teffy}).}
\end{figure}
\begin{figure}[tbp]
\vspace{8.0cm}
\includegraphics{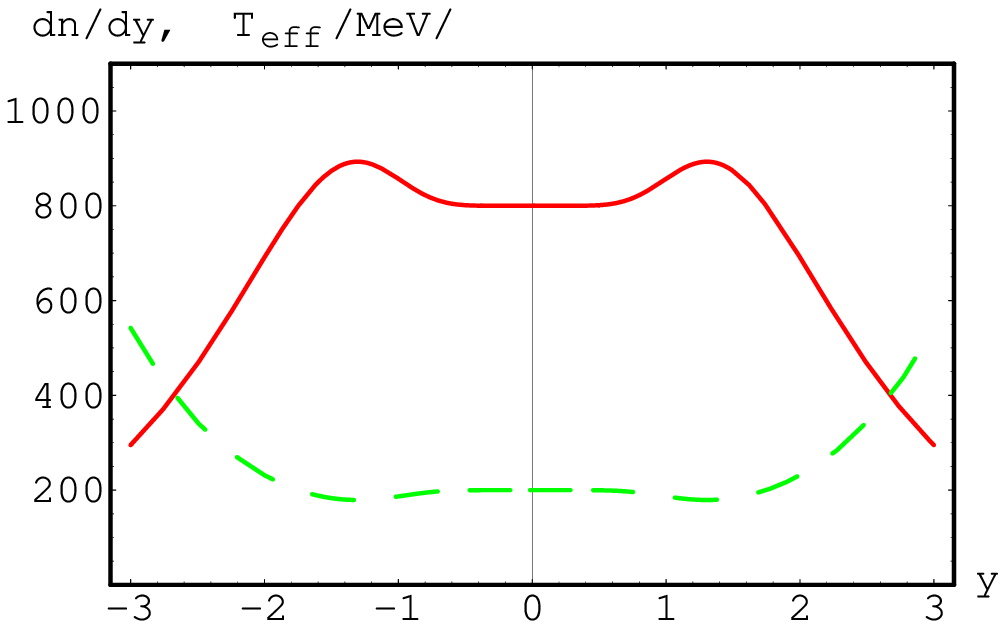}
\vspace{-1.3cm}
\caption{Same as Fig. 1 but utilizing a different  form of the
scaling function, 
$\NN(x) = \sqrt{1 + 1.6  x^4 - 2.6 x^8}$, using 
a scale parameter $\dot Z_0^2 = 1$, $n_0 t_0 = 800$  and $T_0 (t_0/\tau_f)^{1/\kappa} = 200$   MeV,
corresponding to a two-peaked rapidity distribution.}
\end{figure}
\vfill\eject

\begin{thebibliography}{99}
\bibitem{csernai}L. P. Csernai, \textit{Introduction to Relativistic Heavy Ion
Collisions}, John Wiley and Sons, 1994

\bibitem{Shu}D. Teaney, J. Lauret and E.V. Shuryak,  nucl-th/0110037 v1

\bibitem{Hirano}T. Hirano, K. Morita, S. Muroya and C. Nonaka,
nucl-th/0110009 v1

\bibitem{Kolb}P.F. Kolb, U. Heinz, P.Huovinen, K.J. Eskola and K. Tuminen,
hep-ph/0103234 v3, Nucl. Phys. A696 (2001) 197-215

\bibitem{SPH-qm01}C.E. Aguiar, Y. Hama, T. Kodama, T. Osada, Nucl. Phys. A698 (2002) 639-642. 

\bibitem{SPH-jpg}C.E. Aguiar, T. Kodama, T. Osada, Y. Hama, J. Phys.
\textbf{G27} (2001) 75-94

\bibitem{Landau}L.D. Landau, Izv. Akad. Nauk SSSR 17 (1953) 51;  in
``Collected papers of L. D. Landau" (ed. D. Ter-Haar, Pergamon,  Oxford, 1965)
p. 665 - 700

\bibitem{Bjorken}J.D. Bjorken, Phys. Rev. \textbf{D27} (1983) 140

\bibitem{biro1}T.S. Bir\'o, Phys.Lett. B474 (2000) 21-26

\bibitem{biro2}T.S. Bir\'o, Phys.Lett. B487 (2000) 133-139

\bibitem{jnr}J. Bondorf, S. Garpman and J. Zim\'{a}nyi, Nucl. Phys.
\textbf{A296} (1978) 320 .

\bibitem{jde}J.N. De, S.I.A. Garpman, D. Sperber, J.P. Bondorf and J.
Zim\'{a}nyi, Nucl. Phys. \textbf{A305} (1978) 226.

\bibitem{nr}T. Cs\"org\H{o}, B. L\"orstad and J. Zim\'anyi;  Phys. Lett.
\textbf{B338} (1994) 134; nucl-th/9408022

\bibitem{nrt}J. Helgesson, T. Cs\"org\H{o}, M. Asakawa and  B. L\"orstad,
Phys. Rev. \textbf{C56} (1997) 2626.

\bibitem{cspeter}P. Csizmadia, T. Cs\"{o}rg\H{o} and B. Luk\'{a}cs,
nucl-th/9805006, Phys. Lett. B \textbf{443} (1998) 21.

\bibitem{cssol}T. Cs\"{o}rg\H{o}, nucl-th/9809011

\bibitem{ellsol}S.V. Akkelin, T. Cs\"org\H{o}, B. Luk\'acs,  Yu.M. Sinyukov
and M. Weiner, Phys. Lett. B \textbf{505} (2001),  64.

\bibitem{ellsp}T. Cs\"{o}rg\H{o}, S.V. Akkelin, Y. Hama, B. Luk\'{a}cs and
Yu.M. Sinyukov, hep-ph/0108067.

\bibitem{3d}T. Cs\"org\H o and B. L\"orstad, Phys. Rev. \textbf{C54} (1996)
1390; T. Cs\"org\H o and B. L\"orstad, Nucl.  Phys. \textbf{A590} (1995) 465c.
\end{thebibliography}
\end{document}